# PERFORMANCE EVALUATION ON THE BASIS OF BIT ERROR RATE FOR DIFFERENT ORDER OF MODULATION AND DIFFERENT LENGTH OF SUBCHANNELS IN OFDM SYSTEM

Sutanu Ghosh

Department of Electronics and Communication Engineering
Dr Sudhir Chandra Sur Degree Engineering College.
Kolkata, India.

## ABSTRACT

*Today, we have required to accommodate a large number of users under a single base station. This can be possible only if we have some flexibility over the spectrum. Previously we have lots of multiplexing methods to accommodate large number of signals in time and frequency domain. But now we have required to accommodate a large number of users in the same bandwidth, without any fading over the received signal. So, orthogonality can be maintained over the frequency response. This technology is now more popular in the mobile communication domain, called Orthogonal Frequency Division Multiplexing (OFDM). Actually user data can be converted into the parallel form and then they are modulated using digital modulation techniques. Finally, they have followed by OFDM Modulator and cyclic prefix can be inserted into the OFDM symbols. Here, I have worked on the measurement of Bit error rate for different modulation techniques in OFDM technology. It has been considered that subchannel size is not constant. According to that I have concluded the overall idea regarding the performance under OFDM technology.*

## 1. INTRODUCTION

From last two decades communication domain has centered of attention on modulation and multiplexing techniques to provide broadband transmission over wireless channels. OFDM is a multi carrier transmission technique [1] is used to provide high speed data rate over wireless noisy channels at low amount of network complexity. Here the different subcarriers can be modulated separately [2]. It has the ability to reduce the ISI and ICI, even it doesn't require equalizer. ISI can be directly minimized using Cyclic Prefix (CP), which is simply a guard time. This technology is most useful due to it's better spectral efficiency and power efficiency. But OFDM has some disadvantages - high peak to average power ratio (PAPR) [3] and bit error rate (BER). Bit error rate is the important parameter to calculate the end to end performance measurement. There have so many steps to generate this OFDM signal. The OFDM transceiver action can be made on a block by block basis. Here, inverse Fast Fourier Transform (IFFT) and Fast Fourier Transform (FFT) operation can be performed at the transmitter and receiver end, respectively. At the mapping area and serial to parallel converter before IFFT of transmitting end, different modulation techniques can be used to modulate the symbols. The reverse demodulation operation can be performed at the demapping area and parallel to serial converter after FFT of receiving end. The type of modulation can be decided in accordance with the type of information. There have present different order of modulations. In accordance to the modulation order the amount of error in the bit information can be changed. So, I have worked on that issue to find the bit error rate for different modulation orders and the different number of subchannels. This research issue is based on the simulation work on the basis of bit error rate of different modulation techniques.

Before this work, there was little research on performance evaluation of OFDM. Ref. [4] is performed on the basis of symbol error rate. In this research, new technique was proposed to minimize SER of OFDM systems by adjusting the distribution of transmission power among the subcarriers. The performance analysis in Ref. [5] is performed on the basis of error correction coding and interleaving. Here, the research result was capable to provide an average bit and frame error rates and outage probabilities. The work under Ref. [6] was capable to provide the system bit-error rate on the basis of an expression, which takes into an account of both AWGN and "nonlinear noise" effects. These works are not sufficient for the analysis of system performance on the issue of bit error rate for different modulation techniques with different size of subchannels in OFDM system. To the best of my knowledge, this kind of work has not been done for OFDM system. So, I have worked on this issue to execute a performance analysis on the basis of a comparative graphical result.

The remaining parts of this paper are organized as follows: In Sections II and III, I give an overview of the basic principle of OFDM technology and Mathematical representation of OFDM signal, respectively. Section IV presents an OFDM system with the proper mathematical view and the way of insertion of cyclic prefix. The resource allocation method in OFDM technology can be described in Section V. Section VI gives the idea about methods of modulation in OFDM technique. The performance evaluations of my work are presented in Section VII. Finally, I conclude this paper in Section VIII.

## 2. BASIC PRINCIPLE OF OFDM TECHNOLOGY

In a traditional serial data transmission, the symbols can be transmitted sequentially, with the frequency spectrum of each data symbol allowed to occupy the whole available bandwidth [7].

A parallel data transmission system provides the possibilities to solve many of the problems encountered with serial data transmission systems. In case of parallel system, the several sequential streams of data are transmitted simultaneously, so that at any instant of time many data elements are being transmitted. In such kind of system, the available spectrum of an individual data element normally occupies only a small part of the whole bandwidth as described in figure 1.

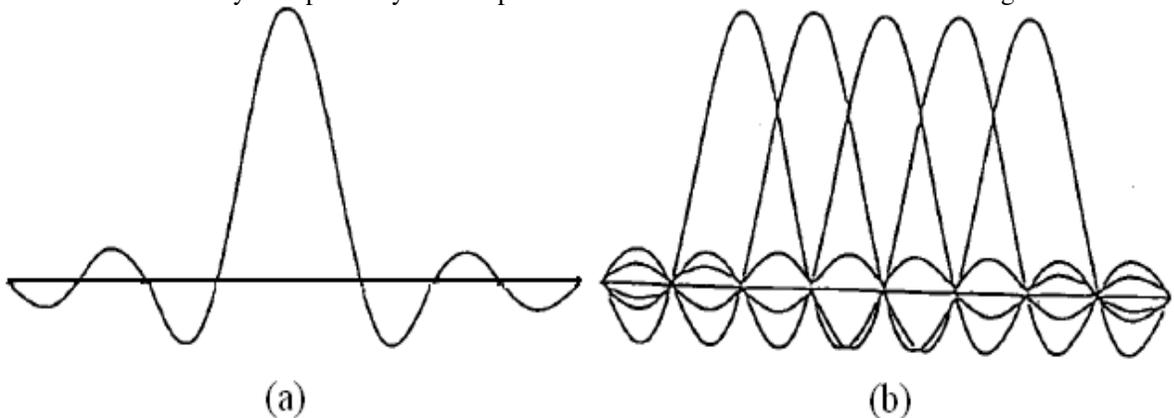

Figure 1: Spectrum of (a) OFDM subchannel and (b) OFDM signal overlapping subcarriers [7]

# I. Mathematical representation of an OFDM signal

From the very beginning of a slot t = $t_s$, the continuous time OFDM symbol can be described as -

$$g(t) = \text{Re}\{\sum_{i=0}^{N-1} b_i \, rect(t - t_s - \frac{T_s}{2}) \exp[\, j2\pi(f_0 + \frac{i}{T_s})(t - t_s)]\} \qquad .....(1)$$

In the above equation t has required to meet $t_s \leq t \leq t_s + T_s$.

$$g(t) = 0 \text{ for } t < t_s \text{ or, } t > t_s + T_s. \qquad ....(1.a)$$

where, N denotes the number of subchannels and $T_s$ denotes the time duration of an OFDM symbols; $b_i$ (i = 0,1,2,...,N-1) is data symbol allocated to each subchannel, $f_0$ is the carrier frequency of first subcarrier; and the function $rect$(t) = $1$, for $|t| \leq T_s/2$ .

The real part is called as in-phase component of the OFDM symbol; and the imaginary part is as called quadrature component.

Now, the output of an OFDM signal using a complex description of equivalent baseband signal can be expressed as:

$$g(t) = \sum_{i=0}^{N-1} b_i \, rect(t - t_s - \frac{T_s}{2}) \exp[\, j2\pi(\frac{i}{T_s})(t - t_s)] \qquad ......(1.b)$$

# II. OFDM System with the insertion of Cyclic Prefix

Figure 3 depicts the overall modulation-demodulation techniques in OFDM system. One end of the transmitter, blocks of information-carrying symbols are converted onto an N number of substreams using serial to parallel converter. Those streams can be modulated using an N number of orthogonal waveforms with frequency $f_k$, where, k= 0,...,N-1. This orthogonal waveform modulation is performed using an IFFT and a parallel to serial converter. The OFDM modulated signal can be computed as the IFFT of the basic smallest elementary unit associated with different subcarriers. The output of the IFFT is considered as the sum of complex exponential functions known as basis functions, complex sinusoids, harmonics, or the tones of a multitone signal [8]. Now, let us consider one of these tones or harmonics, which is the complex exponential function associated with a particular subcarrier. That can be described through the discrete representation:

$$x(n)\,|_{\omega = k\Delta f} = \sum_{k=1}^{N} a_k \, e^{\, j2\pi k \frac{n}{N}} \qquad ....(2)$$

where, $f_k = k/T_s = k\Delta f$ [$\Delta f$ is amount of subcarrier spacing and $T_s$ is symbol duration]; here, $\omega$ is equivalent with $f_k$.

If the channel response is $H_i$ (there are total L+1 number of channels) can be operated on the input transmitted signal then final output can be defined as –

$$y(n) = \sum_{i=0}^{L} H_i x(n - d_i) \qquad ....(3)$$

where, $d_i$ is the amount of multipath delay for the particular assigned channel.

Now, due to linearity issue is concerned, when the OFDM signal is considered for a multipath fading channel then each of its complex exponential components is also subjected to the identical channel model. Therefore, it can be computed that the received version of each subcarrier component of the OFDM signal $(y(n)|_{\omega=k\Delta f})$ as the convolution between transmitted signal and channel impulse response. That can be illustrated by eq. 4

$$y(n)\mid_{\omega=k\Delta f} = \sum_{i=0}^{L} H_i x(n)\mid_{\omega=k\Delta f} \qquad\qquad ....(4)$$

After the converter, last L number of points can be appended to the beginning of the sequence as the CP. This CP is a special kind of spectral time guard in the symbol transitions. Finally the resulting samples are then shaped and transmitted. These transmitted blocks are then referred to as a processed OFDM symbol.

After the addition of CP we can substitute the expression of $x(n)\mid_{\omega=k\Delta f} = a_k e^{j2\pi kn/N}$, if and only if the multipath propagation delay is less than or equal to the length of CP. Otherwise, with even a single delay value outside the range of the length of CP, we cross the OFDM symbol boundary and the orthogonality between the subcarrier components can be lost. Now, it can be assumed that the delay spread is lying within the range of the length of CP. So, received subcarrier component can be defined as a function of transmitted subcarrier –

$$y(n)\mid_{\omega=k\Delta f} = \sum_{i=0}^{L} H_i a_k e^{j2\pi k\frac{(n-d_i)}{N}}$$

$$........(5)$$

Cyclic prefix insertion is an essential function during the generation of OFDM signal. A cyclic prefix is necessary to avoid the interference from previously transmitted OFDM symbols. Cyclic prefix insertion may be observed as a useless operation since it is simply repeats a copy of the existing data in the OFDM symbol and does not add any new information. But it is necessary for multiple reasons. It helps to maintain orthogonality between the subcarriers in the receiver, which is one of the bases of an orthogonal frequency division transmission. This CP is used to provide a periodic extension to the OFDM signal through which a "linear convolution" operation can be performed on the transmitted signal by the channel, can be approximated by a "circular convolution" operation [9].

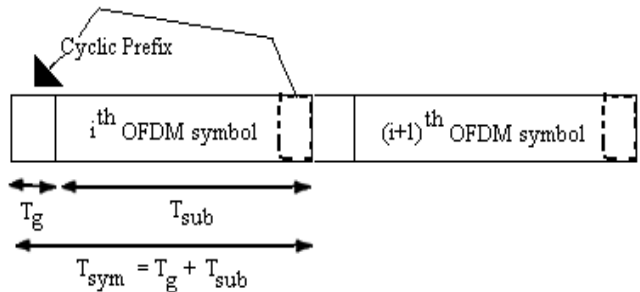

Figure 2: OFDM symbol with CP

Figure 2 depicts the OFDM symbol with the CP. There are two kinds of applicable CP in the present day scenario – normal and extended. $T_{sym}$ is the symbol duration. The mathematical definition of this symbol duration is $-T_{sym} = M/W + T_g$, where, M is the number of samples can be chosen to be the power of 2 and W is the total bandwidth, and $T_g$ is the duration of cyclic prefix.

Now, the receiver can be able to do the reverse operation of transmitter, using a FFT operation. At this end the sampled signals are processed to find their origin point of a block and the proper demodulation window. In the next step, it has required to remove the CP (which contains the ISI) and an N' $(N' = N)$ point sequence is to be converted from serial to parallel form and fed it to the FFT. The outputs of the FFT are symbols modulated on N subcarriers, each multiplied by a complex channel gain. Depending upon the availability of the channel information, different types of demodulation or decoding can be used to recover the information bits. The output of the multipliers are then integrated over the period of 0 to T to get back the estimated signals $A'_{k,0}$ … $A'_{k,N'-1}$, which are then converted from parallel to serial data; after the decoding, binary form of transmitted signal is obtained.

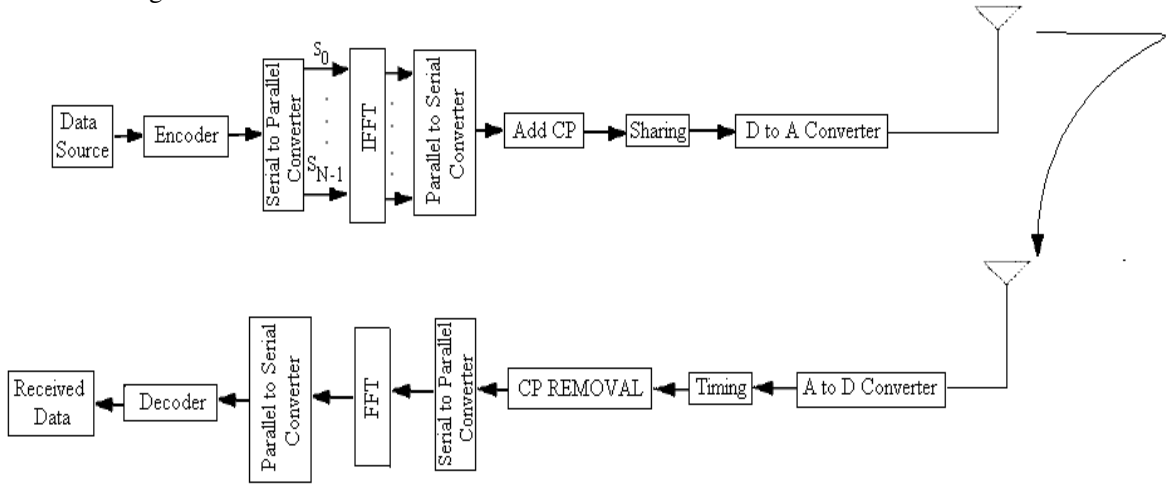

Figure 3: Block diagram of OFDM transceiver system

## 3. RESOURCE ALLOCATION

The traditional fixed resource allocation is not optimal, since the scheme is predetermined regardless of current channel conditions. On the other hand, dynamic resource allocation assigns a dimension adaptively to the users based on their channel gains. Due to the cause of time-varying nature of the wireless channel, dynamic resource allocation makes full utilization of multiuser diversity to achieve higher performance [10]. So, multicarrier application OFDM is most useful to achieve good performance over the other tradition fixed channel application. In case of OFDM access mechanism, a subset of subcarriers is assigned to each user and thus, the number of subcarriers to be assigned to each user must be pre-scheduled by the system. A basic unit of this resource allocation in OFDM access mechanism is subchannel. This subchannel is a group of subcarriers. On the basis of the allocation of subchannel, resource allocation mechanism of OFDM can be categorized in three different classes – Block, Comb and Random type allocation method. Block type is used under the assumption of slow fading channel. Comb Type is used to satisfy the need for equalization, when the channel is varying too fast. Random is used under the consideration of fast fading channel. Block type configuration is used to map the pilot subcarriers on all the subcarriers. Comb type is used to map on the certain number of subcarriers. In case of Random type, pilot subcarrier indexes can be changed periodically [11-12]. There is different length of these subchannels. I have worked with only three different length - 256, 512 and 4096 numbers of subchannels.

## 4. MODULATION TECHNIQUES

In figure 3 the input data stream can be modulated by a QAM modulator and the complex symbol stream can be produced as $S_0$, $S_1$, … $S_{N-1}$. This symbol stream is passed through a serial-to-parallel converter. The output of this converter is a set of N parallel QAM symbols like $S_0$, $S_1$, … $S_{N-1}$ corresponding to the transmitted symbol over each subcarriers [13-14]. At the receiver end the output of the FFT can be passed through the parallel to serial converter and ultimately the output of this converter can be demodulated by the QAM demodulator to recover the original data. Here, I have used different order of modulation with different number of subchannels to observe the bit error rate for OFDM signal. Modulation order has been taken as 4 (QPSK), 8 (8-QAM) and 16 (16-QAM).

# 5. SIMULATION AND EXPERIMENTAL RESULTS

In this research work, 3 different subchannels can be taken to perform a comparative analysis. Here I have done my experimental work through MATLAB simulation software. 100 number of iterations is taken for each of the experimental results. I have taken the following parameters and set their value as given in Table I.

Table I. Set up Values of Experimental parameters

| Parameters | Value |
|---|---|
| Number of Subchannel | 256, 512 and 4096 |
| Total number of pilot | 32 for 256 number of subchannels; 64 for 512 number of subchannels; 512 for 4096 number of subchabnnels |
| SNR Variation (For this issue) | 0 - 27 dB |
| Total number iteration taken for each evaluation | 100 |

**Case study :1**

Number of subchannel taken –256

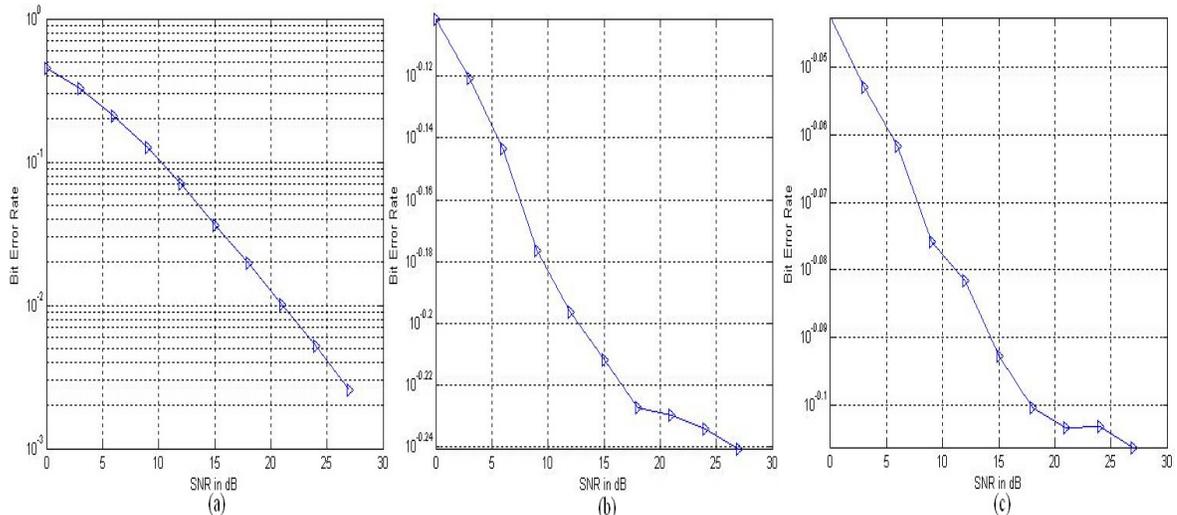

(a)                               (b)                               (c)

Figure 4: Bit error rate for different level of SNR (in dB) for – (a) modulation order 4, (b) modulation order 8, and (c) modulation order 16 with256 number of subchannels

If we observe the above figure then we will get an idea about the amount of bit error rate and best performer (modulation order of 4) within the different order of modulations at 0dB level of SNR, whereas the same kind of result also can be observed at higher level of SNR. The best performer (4th order modulation) has the bit error rate of around 0.00144 (which is the lowest with respect to the other 2 different order of modulations). Finally, I got right option for the higher level of SNR and that is 4[th] order modulation means QPSK.

**Case study :2**

Number of subchannel taken –512

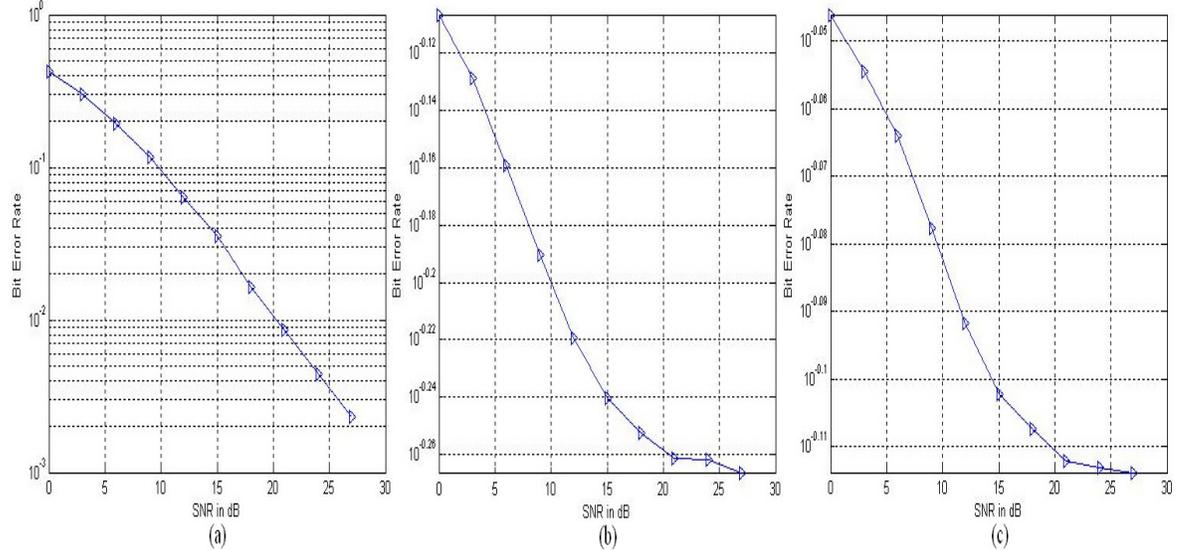

Figure 5 :Bit error rate for different level of SNR (in dB) for – (a) modulation order 4, (b) modulation order 8, and (c) modulation order 16 with 512 number of subchannels

If we compare above two figures (figure 4 and 5), it can be observed that the amount of bit error can be minimized for the two higher order modulation (modulation order 8 and 16) in figure 5 with respect to figure 4, whereas the bit error rate is almost same for lower order modulation (modulation order 4). So, it can be said as, the bit error rate can be influenced by the length of the subchannel.

**Case study :3**

Number of subchannel taken –4096

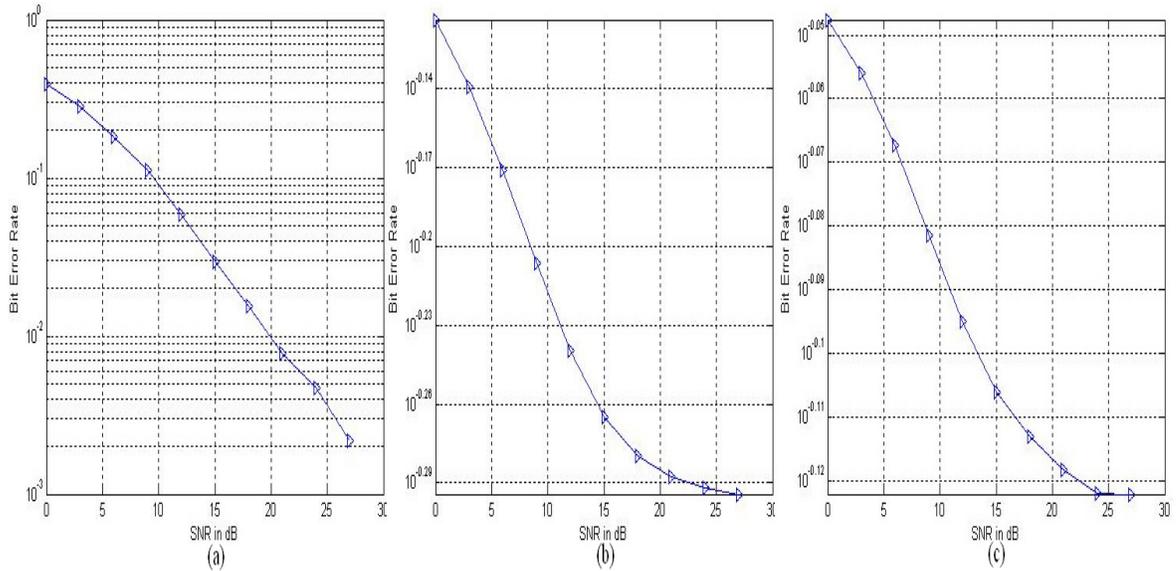

Figure 6 :Bit error rate for different level of SNR (in dB) for – (a) modulation order 4, (b) modulation order 8, and (c) modulation order 16 with 4096 number of subchannels

In this case, I have taken an extra large number of subchannels to observe the continuation of the above result under the case study 2.

**Discussion of the above results**

Finally, I got the expected result, which is similar with the analytical work. If we increase the level of SNR, then the amount of bit error rate for modulation order 16 can be saturated after a certain level of the threshold value.

## 6. CONCLUSION

In this paper, I explore an idea about the comparative performance evaluation on the basis of bit error rate of a different order of modulation with respect to different size of subchannels. I studied that bit error rate can be increased with the increasing size of channel length and bit error rate can be decreased with the increasing size of subchannels. We know that the higher amount of data accommodation can be possible through the higher order modulation. So, it can be concluded that the better performance with higher data accommodation capacity in OFDM system can be available through the large length of subchannel with higher order of modulation.